\documentstyle[preprint,pre,aps,eqsecnum,epsf,graphicx,color]{revtex}
\tightenlines
\newcommand{\beq}{\begin{eqnarray}}
\newcommand{\eeq}{\end{eqnarray}}

\newcommand{\real}{{\sf I}\kern-.12em{\sf R}}
\newcommand{\comp}{{\sf I}\kern-.50em{\sf C}}
\newcommand{\unity}{{\sf I}\kern-.54em{\sf 1}}
\def\spose#1{\hbox to 0pt{#1\hss}}
\def\ltapprox{\mathrel{\spose{\lower 3pt\hbox{$\mathchar"218$}}
\raise 2.0pt\hbox{$\mathchar"13C$}}}

\begin{document}

\rightline{GEF-TH 22/07}
\centerline{\bf Magnetic monopoles in the high temperature phase of
  Yang--Mills theories}
\vskip 5mm
\centerline{A. D'Alessandro and M. D'Elia\footnote{E-mail addresses:
  adales@ge.infn.it, delia@ge.infn.it}} 
\centerline{\it Dipartimento di Fisica, Universit\`a di Genova and
  INFN, Sezione di Genova,}
\centerline{\it Via Dodecaneso 33, I-16146 Genova, Italy}

\begin{abstract}
We investigate the properties of thermal abelian magnetic monopoles
in the high temperature phase of Yang--Mills theories, 
following a recent proposal for their identification on lattice
configurations.
The study is done for SU(2) pure gauge theory, for temperatures
going up to about 10 times the deconfining temperature 
and using the Maximal Abelian gauge to perform the abelian projection.
We find that the monopole density has a well defined continuum limit.
Its temperature dependence disagrees with a free particle gas
prediction and is instead well described by a 
$T^3/(\log (T/\Lambda))^\alpha$ behaviour in all the explored
range, with $\alpha \sim 2$ and $\Lambda \sim 100$ MeV.
Also the study of spatial correlations of thermal monopoles
shows the presence of non-trivial interactions among them.
Finally, we discuss the gauge dependence of our results,
showing that it is significant and that, even within the 
Maximal Abelian gauge, Gribov copy effects are important.
\end{abstract}

\section{Introduction}

Abelian magnetic monopoles are topological defects which may be relevant
for many non-perturbative features of QCD. They enter for instance in 
the mechanism for color confinement based on dual superconductivity of
the vacuum~\cite{thooft75,mandelstam,parisi}, which relates
confinement to the spontaneous breaking of a magnetic symmetry induced
by monopole condensation: the magnetic condensate disappears at the 
deconfining phase transition, as lattice simulations 
have shown extensively~\cite{superI-II,superIII,superfull,superIV,moscow,bari}.

Magnetically charged particles have also been proposed to 
be an important component of deconfined matter above 
the phase transition~\cite{kortals,shuryak,shuryak1,chezak,chelatt}, 
possibly contributing to the physical properties of the 
strongly interacting Quark-Gluon Plasma. 
In Ref.~\cite{chezak} the magnetic
component of the deconfined plasma has been directly related
to thermal abelian monopoles evaporating from the magnetic condensate which is 
present at low temperatures; moreover it has been proposed
to detect such thermal monopoles in 
finite temperature lattice QCD simulations by identifying them 
with monopole currents having a non-trivial wrapping in the Euclidean 
temporal direction~\cite{chezak,bornya92,ejiri}. First numerical investigations
of these wrapping trajectories were performed in Ref.~\cite{bornya92}
and~\cite{ejiri}.

The purpose of this paper is to work on this last proposal
and perform a detailed study of the properties of thermal abelian monopoles.
We will verify if the 
density of such thermal objects is a well defined physical quantity,
i.e. independent of the lattice UV cut-off in the continuum limit, 
determine its temperature dependence, compare it with 
phenomenological models and  study the spatial distribution of the 
wrapping trajectories.
To that
aim we have performed extensive numerical simulations of pure $SU(2)$ 
lattice gauge theory in the deconfined region, in a range of
temperatures going up to $\sim 10\ T_c$.

Abelian monopole currents are exposed on the lattice by the usual
De Grand - Toussaint procedure~\cite{degrand}. That is done on abelian
projected configurations, meaning that the so exposed monopole
currents are gauge dependent quantities. The usual attitude in
the literature is to define and study monopole currents in the gauge 
which maximizes the abelian component of Yang-Mills fields, the
so-called Maximal Abelian gauge (MAG), the rationale being that 
in this way abelian projected fields reproduce most of the original
Yang--Mills dynamics (Abelian Dominance). We have followed that recipe 
in our study. However we believe that the problem of gauge dependence is 
particularly important, especially if one wants to associate the
so detected thermal abelian monopoles with real thermal objects  
influencing the physical properties of the 
Quark-Gluon Plasma: for that reason part of our investigation 
has been dedicated to this issue.

The paper is organized as follows. In Section~\ref{monden} 
we illustrate our numerical simulations and present results
about the density and the 
spatial correlation of wrapping monopole trajectories. 
In Section~\ref{mongau} we investigate the gauge dependence of our results.
Finally, in Section~\ref{discon}, we draw our
conclusions.

\section{Numerical simulations and results}
\label{monden}

We have performed numerical simulations of $SU(2)$ pure gauge theory,
using the standard plaquette action and various lattice sizes
$L_s^3 \times L_t$ at different values of the inverse gauge coupling
$\beta = 2 N_c / g^2$.  The physical scale
has been determined according to
$a(\beta) \Lambda_L = R(\beta) \lambda(\beta)$, where $R$ is the 
two-loop perturbative {$\beta$-function},n while $\lambda$ is a 
non-perturbative correction factor computed and reported in Ref.~\cite{karsch}.
Finally, the values $T_c / \Lambda_L = 21.45(14)$ 
and $T_c / \sqrt{\sigma} = 0.69(2)$ have been assumed respectively 
from Ref.~\cite{karsch} and \cite{karsch2}, where $T_c$ is the
deconfining critical temperature.

Abelian monopole currents are identified by
the De Grand-Toussaint construction~\cite{degrand} after abelian
projection, the last procedure requiring gauge fixing. 
Following the attitude adopted by previous literature, we have 
worked in the Maximal Abelian gauge (MAG), defined by maximizing 
the following functional with respect to gauge transformations:
\begin{equation}
F_{\rm MAG} = \sum_{\mu,x} {\rm Re}\,  \mbox{tr} \left[U_\mu(x)
  \sigma_3 U^{\dagger}_\mu(x) \, 
\sigma_3\right]
\label{maxfun}
\end{equation}
which is proportional to the average squared diagonal part of the
gauge links. It can be shown that on stationary points of the MAG
functional the local operator
\begin{equation}
X(x) = \sum_\mu \left[ U_\mu(x) \sigma_3 U^\dagger_\mu(x)
+U^\dagger_\mu(x-\mu) \sigma_3 U_\mu(x-\mu)\right]
\end{equation}
is diagonal.
The maximization of the MAG functional on a given
configuration has been achieved by an iterative combination of local 
maximization and overrelaxation (see Ref.~\cite{cosmai}), 
stopping the algorithm
when the average squared modulus of the non-diagonal part
of $X(x)$ was less than a given parameter $\omega$~\cite{cosmai}. 
We have chosen $\omega = 10^{-8}$
(see Section~\ref{mongau} for a discussion about the dependence on $\omega$).

After taking the diagonal part of gauge links on the gauge fixed
configuration (abelian projection), monopole currents are defined 
as~\cite{degrand}:
\begin{equation}
m_\mu = {1 \over 2 \pi} \varepsilon_{\mu\nu\rho\sigma} \hat\partial_\nu \overline
\theta_{\rho\sigma}
\end{equation}
where $\overline \theta_{\rho\sigma}$ is the compactified part of the
abelian plaquette phase
\begin{equation}
\theta_{\mu\nu}= \overline \theta_{\mu\nu} + 2 \pi n_{\mu\nu}
\end{equation}
and $n_{\mu\nu}\in \mathbf{N}$.

Monopole currents form closed loops,
since $\hat\partial_\mu m_\mu=0$. These loops may be either
topologically trivial or wrapped around the lattice.
Following the proposal in Refs.~\cite{chezak,bornya92,ejiri}, 
among different currents we select those having a non-trivial 
wrapping around the euclidean temporal direction, identifying them
with thermal monopoles. This procedure
identifies at any given timeslice the number and the spatial positions
of the wrapping trajectories. 
The thermal monopole density is then defined
as~\cite{chezak,bornya92,ejiri}
\begin{equation}
\rho = \frac{\left< \sum_{\vec{x}} \left| N_{wrap}(m_0(\vec{x},t))
\right| \right> }{V_s}
\label{densdef}
\end{equation}
where $N_{wrap}(m_0(\vec{x},t))$ is the temporal winding number 
of the monopole current $m_0$ at site $(\vec{x},t)$, while 
$V_s = (L_s a)^3$ is the spatial volume.

\subsection{Monopole density}

We have performed an extensive study of the thermal monopole density
defined in Eq.~(\ref{densdef}), in a range of temperature going from
$\sim 1.3\ T_c$ to about $10\, T_c$. 
In particular, in order to study the dependence both on the temperature
and on the UV cutoff, we have done 4 different sets of simulations
with parameters $(L_s,\beta)$ = (24, 2.5115), (32, 2.6), (40, 2.7) and
(48, 2.75), corresponding to different lattice spacings but
approximately equal spatial sizes $L_s a(\beta) \sim 2$ fm.
For each set we have performed simulations with different values
of $L_t$ (ranging from 4 to $L_s/4$), 
corresponding to different temperatures $T = 1/(L_t a)$.
A further set of simulations has been done on a $48^3 \times 4$
lattice and inverse gauge couplings up to $\beta = 3.10$, 
in order to extend the range of temperatures explored.
Accumulated statistics range from about 500 decorrelated
configurations for the largest lattices $L_s = 48$ to about 
$5 \cdot 10^3$ for the smallest lattices $L_s = 24$.
All results are reported in 
Table~\ref{denstab}, both in adimensional and in physical units.

\begin{figure}[!htpb]
\begin{center}
\includegraphics*[width=0.81\textwidth]{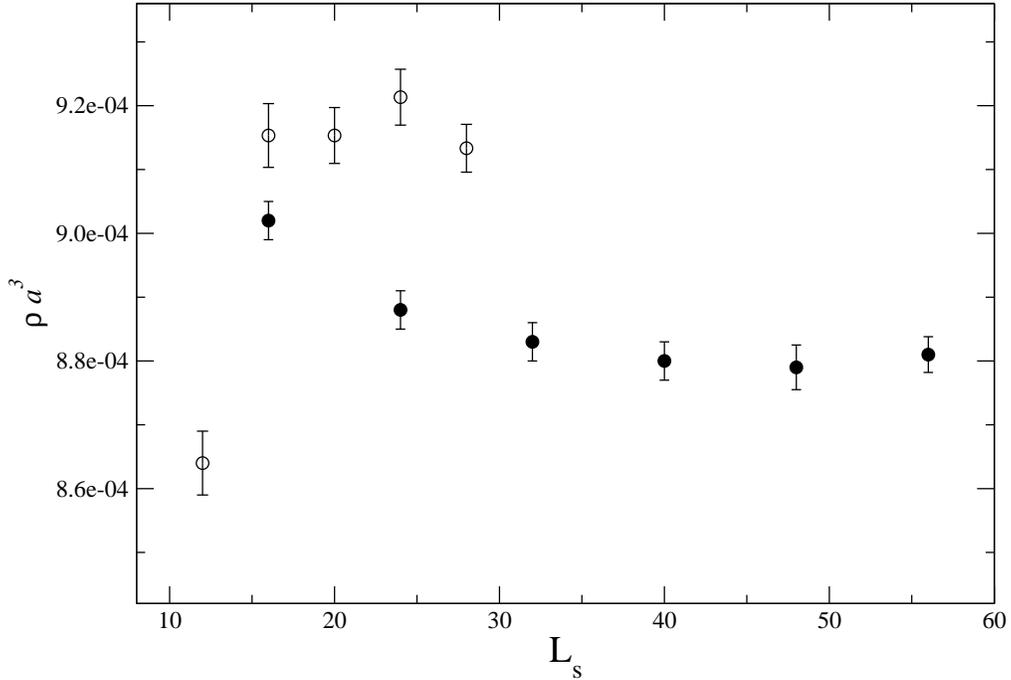}
\caption{Spatial size dependence of the thermal monopole density
 for $\beta = 3.0$ and $L_t = 4$ (filled circles) and for 
 $\beta = 2.5115$ and $L_t = 6$ (blank circles). In the last case 
data have been divided by a factor 1.5 to fit in the figure.
The corresponding data reported in Table~\ref{denstab} have been 
obtained for $L_s = 24$ and
$L_s = 48$ respectively, i.e. well within the two plateau regions.}
\label{Ls3.0}
\end{center}
\end{figure}

We have carefully verified the absence of finite volume effects. 
We show two examples in Fig.~\ref{Ls3.0}, 
where the dependence of $\rho a^3$ on the spatial
size is plotted as a function of $L_s$ for two different combinations
of $\beta$ and $L_t$: in both cases the value
of $L_s$ used to extract our data is well within an extended 
plateau region. A similar check has been performed for the dependence
on the gauge fixing stopping parameter $\omega$: a detailed
discussion about gauge dependence is reported in Section~\ref{mongau}.
We have also checked the absence, in the whole temperature range explored, 
of trajectories wrapping in a spatial instead than in the temporal
direction.

\begin{figure}[h!]
\begin{center}
\includegraphics*[width=0.81\textwidth]{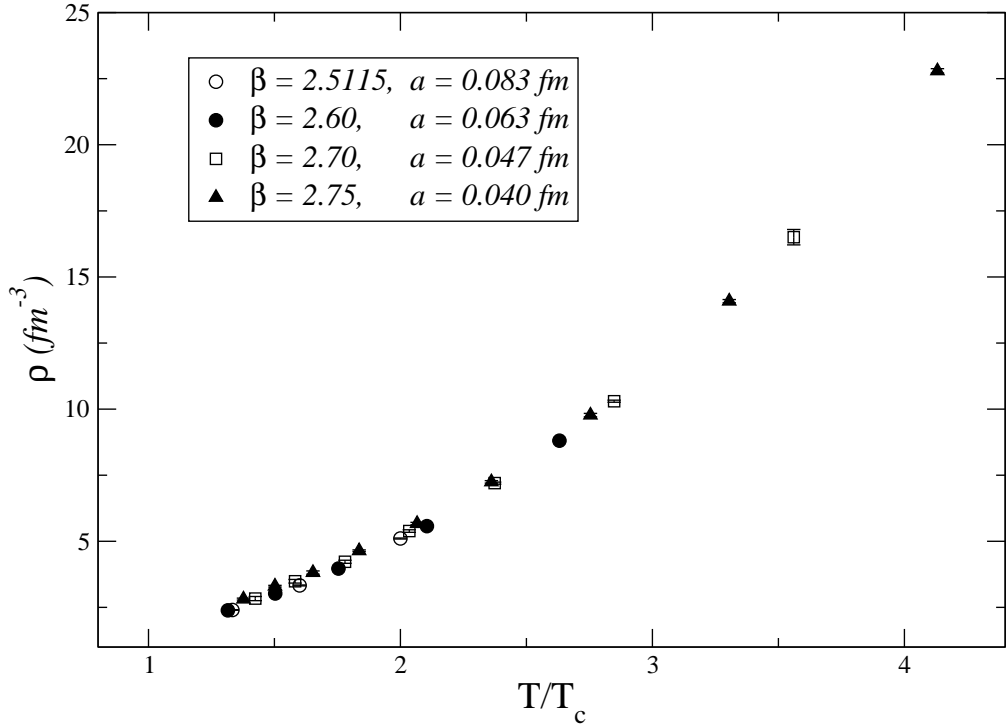}
\vspace{-0.cm}
\caption{Monopole density $\rho(T)$ in ${\rm fm}^{-3}$ for
different values of $T$ and different lattice spacings. 
}
\label{densfig} 
\vspace{-0.cm}
\end{center}
\end{figure}

Let us now discuss the dependence of our results on the UV cut-off.
We report in Fig.~\ref{densfig} results obtained for $\rho$ versus
$T/T_c$ for 4 different values of the lattice spacing. 
The physical scale has been determined as discussed above and 
fixing $\sqrt{\sigma} = 430$ MeV. The agreement among data obtained
at different UV cut-off is reasonable, indicating that the density
of wrapping monopole trajectories has a well defined continuum limit.

\begin{figure}[h!]
\begin{center}
\includegraphics*[width=0.85\textwidth]{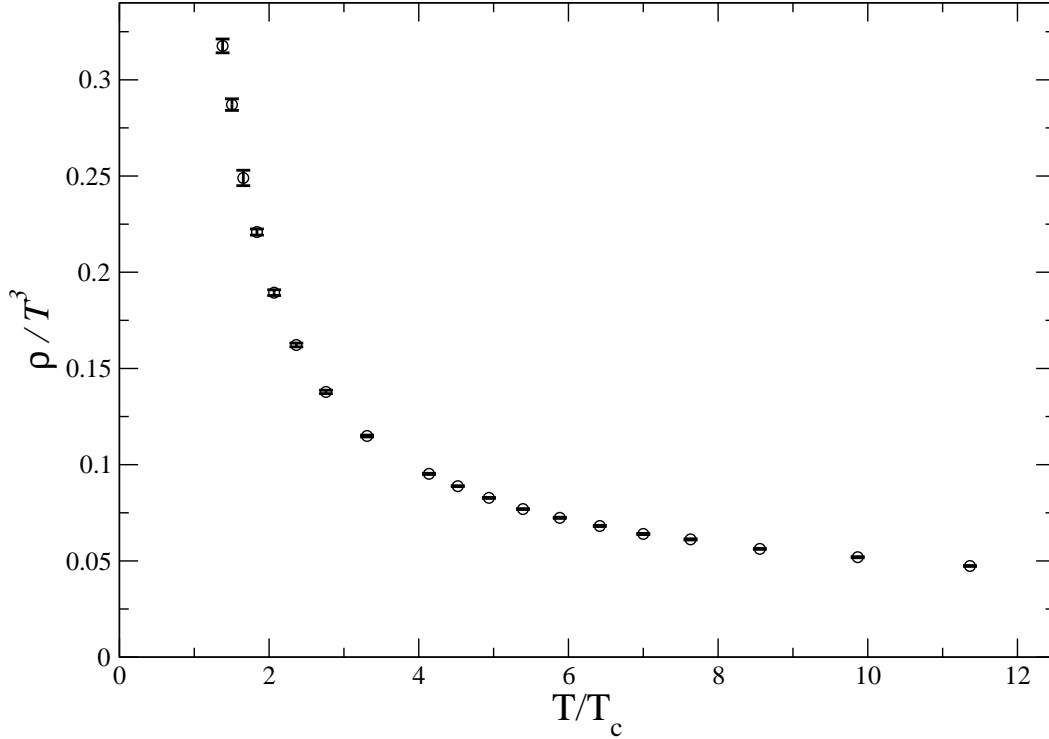}
\vspace{-0.cm}
\caption{$\rho(T)/T^3$ as a function of $T/T_c$. Data have been
  obtained on a $48^3 \times L_t$ lattice, with variable $L_t$ and
  at $\beta = 2.75$ (first 9 points), and variable $\beta$ at $L_t =
  4$ (last 10 points).}
\label{figrhoT3} 
\vspace{-0.cm}
\end{center}
\end{figure}

Next we turn to the temperature dependence of $\rho$. 
The thermal distribution for a gas of free ultrarelativistic
(massless) scalar particles predicts a density: 
\begin{equation}
\int \frac{d^3\vec{p}}{(2\pi)^3} \frac{1}{\exp(\left|\vec
p\right|/T)-1}= \frac{\zeta(3)}{\pi^2}  T^3\, .
\label{relnointeract}
\end{equation}
The behaviour $\rho(T) \propto T^3$
is expected anyway in the $T \to \infty$ limit if the
monopoles can be considered as asymptotically free in that regime.  
We report in Fig.~\ref{figrhoT3} the adimensional ratio
$\rho/T^3 = \rho (a L_t)^3$ versus $T/T_c$, for the data on our 
biggest spatial size ($L_s = 48$) and $\beta \geq  2.75$; notice that $\rho/T^3$ is not
influenced by the systematic error related to the determination of the 
physical scale, which only affects $T/T_c$, hence the horizontal
scale of Fig.~\ref{figrhoT3}. It is apparent from the figure that
the behaviour $\rho \propto T^3$ is not verified in all the explored 
range, i.e. for temperatures up to $T \sim 10\ T_c$. It can be easily 
realized that the deviation from the simple cubic behaviour cannot be 
simply explained by the presence of a finite monopole mass $m_M$, 
i.e. by correcting $\left| \vec p \right| \to 
\sqrt{\left| \vec p \right|^2+m_M^2}$ in the expression of the 
free particle energy in
Eq.~(\ref{relnointeract}): indeed $\rho/T^3$ would be an increasing 
function of $T$ in that case, in clear contrast with the behaviour
visible in Fig.~\ref{figrhoT3}.

We conclude that the description 
of thermal monopoles as a gas of free particles is not appropriate,
and interactions must be taken into account also at very high
temperatures. That is in agreement with the scenario of an electric
dominated phase for Yang--Mills theories at very high temperatures,
in which the magnetic component is strongly
interacting~\cite{shuryak}.
Predictions have been made for the density of magnetic 
objects~\cite{giovannangeli},
based on perturbative and dimensional reduction considerations, 
leading to a behaviour like $g^6\ T^3$, hence a reduction factor
with respect to the free massless particle case of the order of
$1/(\log (T/\Lambda))^3$; similar predictions have been reported in 
Ref.\cite{shuryak}.

Based on that, we have tried to correct the cubic behaviour as follows:
\begin{equation}
\frac{\rho}{T^3} = \frac{A}{(\log (T/\Lambda_{eff}))^\alpha}
\label{eqlogt}
\end{equation}
and we have verified that this simple correction describes very well our
data in all the explored range. In particular, a fit including 
data with $T > 2\ T_c$ leads
to $A = 0.48(4)$, $T_c/\Lambda_{eff} = 2.48(3)$ and $\alpha =
1.89(6)$, with $\chi^2/{\rm d.o.f.} = 10.6/12$. Results do not
change much if the range of fitted points is changed, with $\alpha$
always compatible or marginally compatible with $2$ and 
$\chi^2/{\rm d.o.f.}$ of order 1. If we fix $\alpha = 2$
we obtain $A = 0.557(10)$, $T_c/\Lambda_{eff} = 2.69(3)$ (i.e. 
$\Lambda_{eff} \sim 110$ MeV) and
$\chi^2/{\rm d.o.f.} = 13.7/13$ for data corresponding to $T > 2\
T_c$; results are completely stable within errors if the fitted 
temperature range is changed, with only a slight increase in 
$\chi^2/{\rm d.o.f.}$ (up to $23/17$) if temperatures down to $\sim
1.4\ T_c$ are included. If we instead fix $\alpha = 3$, reasonable
fits are obtained for $T > 5\ T_c$, while $\alpha = 1$ seems to
be excluded by our data.

\begin{figure}[h!]
\begin{center}
\includegraphics*[width=0.85\textwidth]{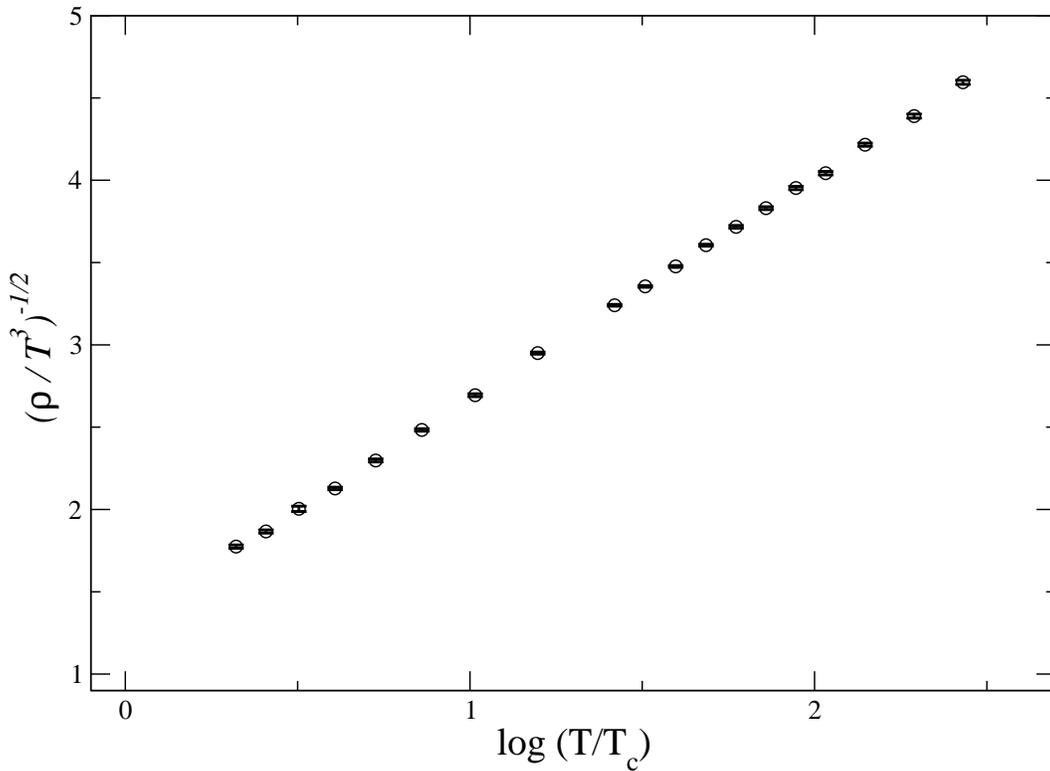}
\vspace{-0.cm}
\caption{$\sqrt{(\rho(T)/T^3}$ versus $\log (T/T_c)$. The data are
  the same reported in Fig.~\ref{figrhoT3}. 
The linear dependence is manifest.}
\label{figrhologT} 
\vspace{-0.cm}
\end{center}
\end{figure}

We conclude that the simple ansatz in Eq.~(\ref{eqlogt}) perfectly
describes our data, with $\alpha \sim 2$, for all temperatures ranging 
from $T \sim 2\ T_c$ up to $T \sim 10\ T_c$ and  with $\Lambda_{eff}$
of the order of $\Lambda_{\rm QCD}$. In order to better 
appreciate that, we have plotted in Fig.~\ref{figrhologT}
the quantity $(\rho/T^3)^{-1/2}$ versus $\log (T/T_c)$: the dependence
is manifestly linear, as Eq.~(\ref{eqlogt}) predicts in the case  $\alpha = 2$.

\subsection{Monopole interactions}
\label{monint}

In order to further investigate the interactions existing
among wrapped monopole trajectories, we have studied their 
distribution at a fixed time slice, so as to obtain information
about the spatial correlation of thermal monopoles.
In particular we have determined the density--density correlation function
$g(r) \equiv \langle \varrho(0) \varrho(r) \rangle/ \varrho^2$, 
which can also be
measured (for $r \neq 0$) as the ratio between the probability  
of having a particle at a distance $r$ 
from a given reference particle and the 
same probability in a completely homogeneous system, 
in practice
\begin{equation}
g(r) = \frac{1}{\varrho} \frac{d N(r) }{4 \pi r^2 dr}
\end{equation}
where  $d N (r)$ is the number of particles in a spherical shell
of thickness $d r$ at distance $r$ from the reference particle and by
$\varrho$ we mean the average density of such particles (monopoles
or antimonopoles).
A value $g(r)<1$ ($g(r)>1$) indicates that at distance $r$ we have less (more)
particles than expected in a non-interacting medium, i.e. there is a
repulsive (attractive) interaction. In our numerical
study we have used, in place of $4 \pi r^2 dr$, the actual number
of lattice sites contained  in the shell, in order to take into
account part of the discretization effects.

\begin{figure}[!htpb]
\begin{center}
\includegraphics*[width=0.85\textwidth]{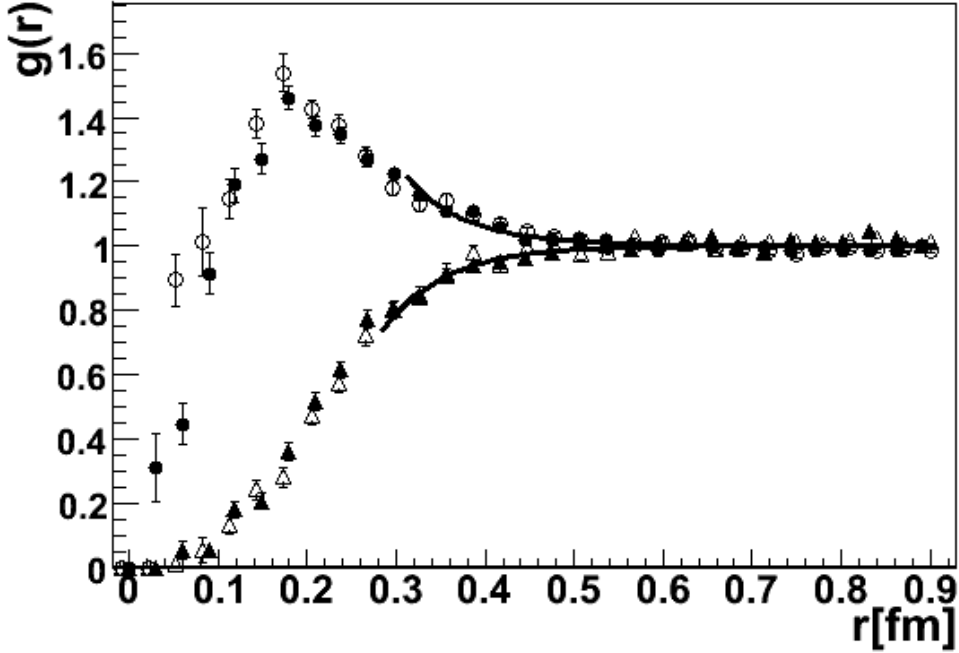}
\caption{$g(r)$ for the monopole-monopole (triangles) and 
the monopole-antimonopole (circles) case and for two different
lattice sizes and $\beta$ values corresponding to 
the same physical volume and temperature ($T \simeq 2.85\ T_c$): 
$40^3 \times 5$ at $\beta=2.7$ (empty markers) 
and $64^3 \times 8$ at $\beta = 2.86$ (full markers). 
The reported
curves correspond to fits according to $g(r) = \exp (-V(r)/T)$
with $V(r)$ a Yukawa potential (see Eqs.~(\ref{yuk1}) and (\ref{yuk2})).}
\label{attrrep}
\end{center}
\end{figure}

We have measured $g(r)$ for both the monopole-monopole and 
the monopole-antimonopole case in some of our simulations.
In Fig.~\ref{attrrep} we show the results obtained at 
$\beta = 2.7$ on a lattice with $L_s = 40$ and $L_t = 5$
($T \simeq 2.85\ T_c$) on a set of 300 decorrelated configurations.
$g(r)$ is clearly depleted at short distances
in the monopole-monopole case, indicating a repulsive
interaction. On the contrary $g(r)$ is enhanced
at short distance in the monopole-antimonopole case, indicating
an attractive interaction. 
Actually in this last case (monopole-antimonopole)
$g(r)$ has a peak at distances of about 
$0.2$ fm (4 lattice spacings) and then goes down again, reaching
values $<\ 1$: 
that could either be related to finite lattice spacing 
effects or have a physical meaning.

This is an important issue that must be clarified, therefore
we have decided to repeat the measurements on a lattice with 
$L_s = 64$ and $L_t = 8$ at $\beta = 2.86$: 
we have $a(\beta = 2.86)/a(\beta = 2.7) \simeq 5/8$
within a good precision, hence the new measurements correspond to
equal physical volume and temperature but different UV cutoff.
Also in this case we have collected a set of 300 decorrelated 
configurations. Results are reported again in Fig.~\ref{attrrep},
the distance $r$ is reported in physical units so that 
comparison is straightforward.
A nice scaling to the continuum
limit can be appreciated, apart from results obtained at a distance of
$\sim 1$ lattice spacing.
In particular we can confirm that the peak observed in the 
monopole-antimonopole correlation at $\sim\ 0.2$ fm is not a lattice
artifact but the signal of more structured interaction patterns. 

\begin{figure}[!htpb]
\begin{center}
\includegraphics*[width=0.75\textwidth]{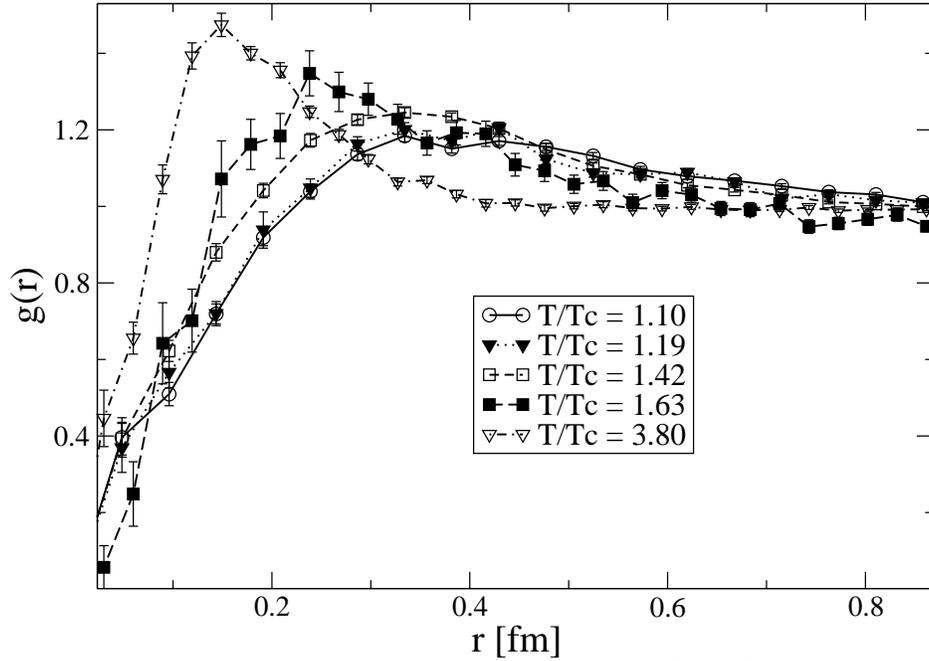}
\caption{
$g(r)$ in the monopole-antimonopole case determined for different
values of the temperature $T$.}
\label{attrrep2}
\end{center}
\end{figure}

In Fig.~\ref{attrrep2} we illustrate some results regarding the temperature
dependence of monopole interactions. The function $g(r)$ shows
appreciable changes as $T$ is varied, even if not dramatic from a 
qualitative point of view. In particular the peak in the
monopole-antimonopole correlation function moves to larger distances
as $T$ is lowered.

In the large distance region, where $g(r) \simeq 1$, one can try to 
extract direct information about the interaction potential $V(r)$ based on the ansatz 
(and assuming $V(\infty) = 0$):
\begin{equation}
g(r) \simeq \exp(-V(r)/T) 
\label{yuk1}
\end{equation}
We have tried this strategy, finding that $V(r)$ can be described reasonably 
well by a Yukawa potential 
(two sample fits are reported in Fig.~\ref{attrrep}) 
\begin{equation}
V(r) \propto \frac{e^{-\lambda_P r}}{r}
\label{yuk2}
\end{equation}
with a plasma screening length $\lambda_P$ of the order
of 0.1 fm. 

\section{Gauge dependence}
\label{mongau}

\begin{figure}[!htpb]
\begin{center}
\includegraphics*[width=0.85\textwidth]{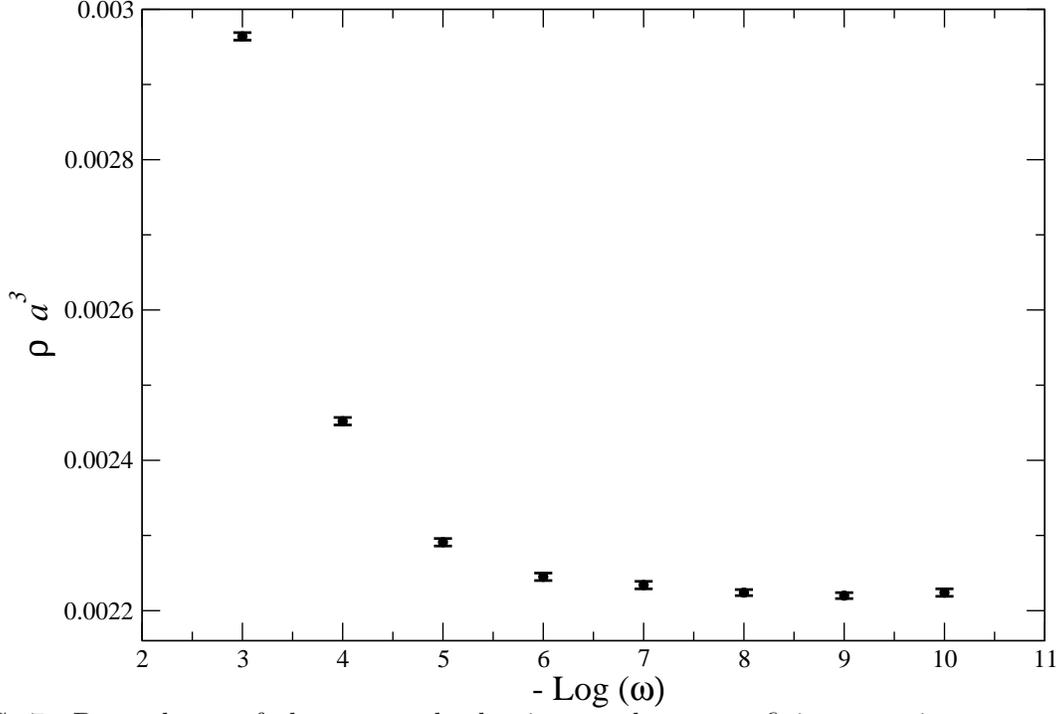}
\caption{Dependence of the monopole density on the gauge fixing
stopping parameter for  $L_s = 32$, $L_t = 4$ and $\beta = 2.6$.
}
\label{gaugedep}
\end{center}
\end{figure}

We now discuss the issue of gauge dependence. The fact 
that monopole currents identified by abelian projection
are gauge dependent is well known: much likely that applies also to
monopole currents wrapping in the temporal direction.
A way to detect this phenomenon is to study the dependence
of the measured thermal monopole density $\rho$ on
the gauge fixing stopping parameter $\omega$:
larger values of $\omega$ correspond to a looser gauge fixing. 
An example of such study is reported in Fig.~\ref{gaugedep},
corresponding to $L_s = 32$, $L_t = 4$ and $\beta = 2.6$.
When the gauge is poorly fixed
the monopole density is larger: a possible interpretation
could be that much more ultraviolet noise contributes
to the monopole currents in this case, whose effect is that of mimicking 
in some case the presence of additional wrapping trajectories.
However the density is well stable for $\omega < 10^{-7}$. 

In all the simulations discussed above we have taken $\omega=10^{-8}$, if a
looser criterion is chosen then the continuum scaling of the monopole
density, which is approximately verified on our data shown in
Fig.~\ref{densfig}, is lost. To show an example of that, we have reported 
in Table~\ref{denstab} two values of the density $\rho$, signalled by a star in the last
column, obtained with a stopping criterion $\omega = 10^{-3}$
on two lattices with different UV cutoff but
approximately equal physical temperatures: quite different value are
obtained for $\rho$, contrary to what obtained by fixing the gauge
more carefully.

What we have shown seems to indicate that thermal monopoles
populating the Quark-Gluon Plasma can be indeed identified in lattice
simulations and that their density is a well defined
physical quantity
as soon as a careful gauge fixing is performed.
However one may wonder how results change if a different gauge is
chosen, in place of MAG, to perform the Abelian projection.
The answer is that results change in a dramatic way: an extreme
case is provided by Landau gauge, which on the lattice is defined by 
maximizing the following functional
\begin{equation}
F_L = \sum_{\mu,x} {\rm Re}\, \mbox{tr}\ U_\mu(x) \, .
\label{Landaufunc}
\end{equation}
We have repeated part of our measurements in Landau gauge, obtaining
exactly zero wrapping monopole trajectories in all the explored cases.
That may sound as a warning, nevertheless 
one could still suppose that the Maximal Abelian gauge has the 
virtue of correctly identifying the magnetic component of the
Yang--Mills plasma in the form abelian magnetic monopoles.
We have however a further warning.

\begin{figure}[!htpb]
\begin{center}
\includegraphics*[width=0.85\textwidth]{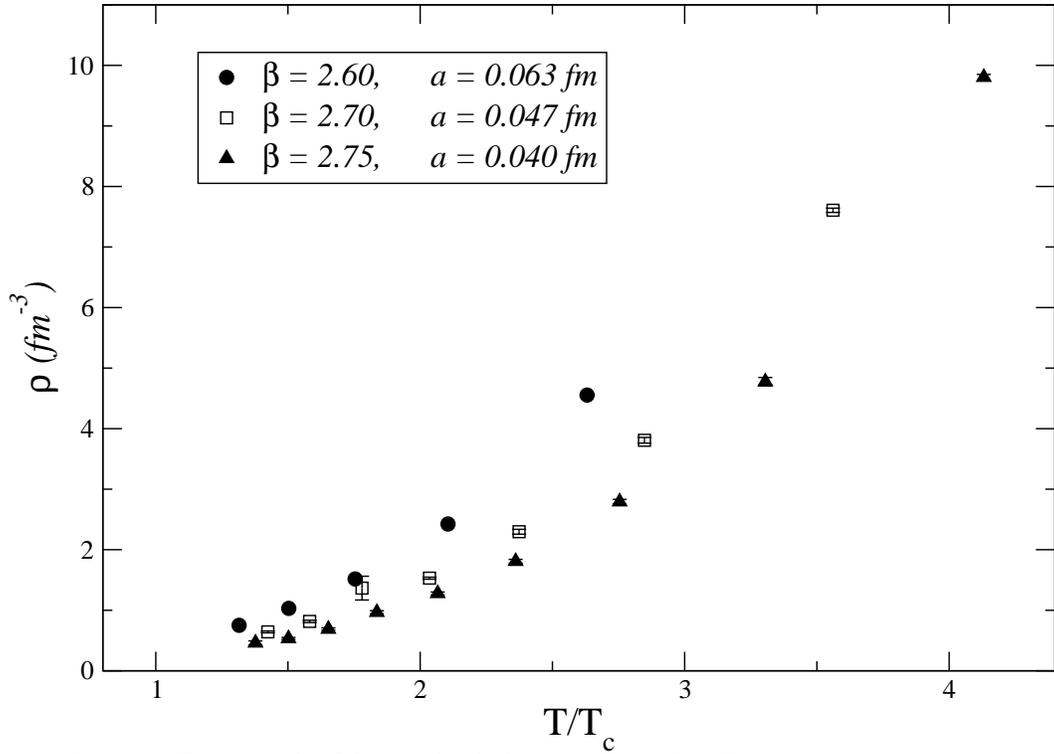}
\caption{Same as Fig.~\ref{densfig} if the Maximal Abelian gauge is
  fixed starting from a maximum of the Landau functional.
}
\label{LMAGscal}
\end{center}
\end{figure}

We have tried the following experiment, which is not new and has been 
performed previously in the literature, even if mostly in the context 
of center dominance studies~\cite{kovacs,bornyakov,greensite,su3gribov}. 
We have repeated part of our measurements, 
using again the Maximal Abelian gauge in order to define 
the abelian projection, but using a sample of configurations
which had been previously fixed to Landau gauge.
The result is that thermal abelian monopoles, 
which are completely absent in Landau
gauge, do reappear when fixing again to MAG. However the average
monopole density is appreciably lower and, what is worse, scaling to
the continuum limit is lost. All that is apparent from Fig.~\ref{LMAGscal}.
What is more striking is that gauge configurations which are 
fixed to MAG after Landau gauge fixing (and which show no
continuum scaling) reach on average higher values of the MAG
functional as compared to original configurations (where
instead scaling is observed), as shown for 
a few examples in Table~\ref{MAGfunctab}. 

All that exposes quite clearly the problem of Gribov copies. While looking
for the global maximum of the MAG functional in Eq.~(\ref{maxfun}) along the 
gauge orbit of a given configuration, several local maxima are met
where a gauge fixing algorithm like the one used by us may stop. 
All these local maxima have a different content of thermal abelian
monopoles, but on average they furnish a monopole density that scales 
correctly in the continuum limit. 
If however one starts the same algorithm from 
points of the gauge orbit corresponding to local maxima of the 
Landau functional, one reaches local maxima of the MAG functional 
that, most of the times, are closest in value to the global maximum but have
a quite different monopole content which, if continuum scaling is taken 
as a criterion, one would say to be wrong.

As we have emphasized above, very similar problems are met when
studying center dominance in the maximal center gauge
(MCG)~\cite{kovacs,bornyakov,greensite,su3gribov} 
and some possible
interpretations have been discussed~\cite{greensite,zak2}.
One should also study how the problem is related to the choice of 
a particular starting point on the gauge orbit (belonging to Landau
gauge in our case) and perform more extensive studies in which
the global maximum of the MAG functional is searched for via better
algorithms, like for instance simulated annealing~\cite{bornya02}.
Our opinion is that much has still to be clarified in this context and
that the issue of gauge dependence is even more urgent in this case,
if one wants to associate wrapping monopole trajectories 
with real thermal objects populating the Quark-Gluon Plasma.

\section{Conclusions}
\label{discon}

Magnetic monopoles may be an important component of the Quark-Gluon
Plasma~\cite{kortals,shuryak,shuryak1,chezak,chelatt}.
In the present
work we have followed a recent proposal aimed at identifying the magnetic
component of the high temperature phase of Yang-Mills theories with 
thermal abelian monopoles detected on
lattice configurations in terms of monopole trajectories wrapping in the
euclidean temporal direction~\cite{chezak,bornya92,ejiri}: 
our purpose has been that of investigating
the temperature and UV cut-off dependence of the density of thermal
monopoles and to study their interactions in the case of $SU(2)$ pure
gauge theory. Abelian monopole currents have been identified in the 
Maximal Abelian gauge, therefore part of our efforts have been dedicated
to the issue of gauge dependence: we have shown that it is indeed 
significant and our physical results should be considered in that light.

We have determined the monopole density $\rho$ in a wide range of
temperatures, going from $\sim 1.3\ T_c$ to $\sim 10\ T_c$, and 
for different values of the lattice spacing. We have verified 
a reasonable scaling of $\rho$ to the continuum limit.
Regarding the temperature dependence, our data show 
that $\rho$ is not compatible with a (massive or massless) free
particle behaviour in the whole range of temperatures explored.
This is in agreement with the picture of an electric dominated
phase for Yang--Mills theories at very high temperatures, in which 
the magnetic component is strongly interacting~\cite{shuryak}. 
Inspired by that picture and by predictions based on perturbative
computations and dimensional reduction~\cite{giovannangeli}, 
we have tried to fit our data according
to $\rho \propto T^3/(\log T/\Lambda_{eff})^\alpha$, finding 
that a coefficient $\alpha = 2$ and $\Lambda_{eff} \sim 100$ MeV 
describe perfectly our data in all the explored
temperature range. A coefficient $\alpha = 3$ is not excluded
for $T > 5\ T_c$. 

By studying the distribution of wrapping monopole trajectories at a given
timeslice, we have extracted information about density--density
spatial correlation functions both in the monopole--monopole and in the
monopole-antimonopole case. We have thus verified the presence of 
a repulsive (attractive) interaction at large distances,  
in agreement 
with a screened Coulomb potential and a screening length of the order
of 0.1 fm. Repulsion is observed at short distances in both cases,
with a peak in the monopole-antimonopole correlation function
at a distance which we have verified to scale well to the continuum
limit and which is about $0.15$ fm at $T \simeq 3.85\ T_c$ and 
goes up to $\sim 0.4$ fm close to the critical temperature.
More extensive future studies could better clarify
the nature of the interaction, both in the high $T$ region and
closer to $T_c$, where one could search for a possible 
interplay with the dynamics of the deconfinement transition.
Of course the extension of our results to $SU(3)$ pure gauge theory 
will be the natural continuation of our study.

Let us stress once again that all results apply to thermal
abelian monopoles identified in a particular gauge. 
As we have shown in our study 
presented in Section~\ref{mongau}, 
results are strongly gauge dependent, an extreme case being
Landau gauge, where thermal abelian monopoles completely
dissappear.
This dependence is well known and the Maximal Abelian gauge 
has been always considered
as a special gauge in which  abelian monopoles are better identified.
Moreover, we have shown that Gribov copy effects may change results
even within a given gauge, if particular points are chosen 
along the gauge orbit where the gauge fixing procedure is started.
We have considered the example of the Maximal Abelian gauge fixed 
starting from local maxima of the Landau functional: higher maxima 
of the MAG functional are reached but the continuum scaling of the
monopole density is lost. One should study how this problem is related
to the particular starting point on the gauge orbit and perform more 
extensive studies, looking for the global maximum of the MAG functional 
via dedicated algorithms like simulated annealing~\cite{bornya02}.

In conclusion, our opinion is that if one really wants to associate 
thermal abelian monopoles with (part of) the magnetic component of the 
Quark-Gluon Plasma, the issue of gauge dependence should be better
clarified: more numerical and theoretical efforts have to be 
done in this respect.

\section*{Acknowledgments}

We thank C.~M.~Becchi, V.~Bornyakov, M.~Chernodub, A.~Di~Giacomo, 
C.~Korthals-Altes, E.~Shuryak and  V.~Zakharov for many interesting 
discussions. We would also like to thank A. Brunengo and M. Corosu for
their kind technical support in the use of the PC farm at INFN - Genova, where
numerical simulations have been performed.

\begin{table}
\begin{center}
\begin{tabular}{|c|c|r|l|r|l|}
\hline
$\beta$  &  $L_s$  & $ L_t$ & $<N_{wrap}>/L_s^3$ & $T/T_c$
 & $\rho 
({\rm fm}^{-3})$ \\
\hline 2.5115 & 24 & 6 & 0.001384(7)   & 1.333   &  2.404(12) \\
\hline 2.5115 & 24 & 5 & 0.001916(10)  & 1.600   &  3.331(18) \\
\hline 2.5115 & 24 & 4 & 0.002938(10)  & 2.000   &  5.107(19) \\
\hline 2.6    & 32 & 8 & 0.000604(4)   & 1.315   &  2.387(15) \\
\hline 2.6    & 32 & 7 & 0.000765(4)   & 1.503   &  3.025(16) \\
\hline 2.6    & 32 & 6 & 0.001003(3)   & 1.754   &  3.969(13) \\
\hline 2.6    & 32 & 6 & 0.001550(5)   & 1.754   &  6.135(20)  *\\
\hline 2.6    & 32 & 5 & 0.001408(3)   & 2.105   &  5.574(13) \\
\hline 2.6    & 32 & 4 & 0.002223(4)   & 2.631   &  8.806(25) \\
\hline 2.7    & 40 & 10& 0.000289(8)   & 1.424   &  2.834(8) \\
\hline 2.7    & 40 & 9 & 0.000356(7)   & 1.582   &  3.49(7) \\
\hline 2.7    & 40 & 8 & 0.000431(6)   & 1.780   &  4.23(6) \\
\hline 2.7    & 40 & 8 & 0.000784(15)  & 1.780   &  7.69(15) \, * \\
\hline 2.7    & 40 & 7 & 0.000549(5)   & 2.035   &  5.39(5) \\
\hline 2.7    & 40 & 6 & 0.000734(4)   & 2.374   &  7.20(4) \\
\hline 2.7    & 40 & 5 & 0.001050(4)   & 2.848   &  10.30(4) \\
\hline 2.7    & 40 & 4 & 0.0016818(29) & 3.561   &  16.50(29) \\
\hline 2.75   & 48 & 12& 0.0001838(23) & 1.377   &  2.82(4) \\
\hline 2.75   & 48 & 11& 0.0002157(21) & 1.502   &  3.30(3) \\
\hline 2.75   & 48 & 10& 0.000249(4)   & 1.653   &  3.8(6) \\
\hline 2.75   & 48 & 9 & 0.0003030(22) & 1.836   &  4.64(3) \\
\hline 2.75   & 48 & 8 & 0.000370(3)   & 2.066   &  5.67(4) \\
\hline 2.75   & 48 & 7 & 0.000473(3)   & 2.361   &  7.25(4) \\
\hline 2.75   & 48 & 6 & 0.000638(4)   & 2.754   &  9.77(6) \\
\hline 2.75   & 48 & 5 & 0.000919(4)   & 3.305   &  14.08(6) \\
\hline 2.75   & 48 & 4 & 0.001488(5)   & 4.131   &  22.80(8) \\
\hline 2.78   & 48 & 4 & 0.001388(4)   & 4.521   &  27.87(8) \\
\hline 2.81   & 48 & 4 & 0.001293(4)   & 4.938   &  33.83(10) \\
\hline 2.84   & 48 & 4 & 0.001202(4)   & 5.392   &  40.95(13) \\
\hline 2.87   & 48 & 4 & 0.001131(5)   & 5.884   &  50.07(24) \\
\hline 2.90   & 48 & 4 & 0.001065(5)   & 6.419   &  61.2(3) \\
\hline 2.93   & 48 & 4 & 0.001000(5)   & 7.000   &  74.5(4) \\
\hline 2.96   & 48 & 4 & 0.000956(5)   & 7.631   &  92.3(4) \\
\hline 3.00   & 48 & 4 & 0.000879(4)   & 8.557   &  119.7(5) \\
\hline 3.05   & 48 & 4 & 0.000811(5)   & 9.865   &  169.2(11) \\
\hline 3.10   & 48 & 4 & 0.000740(4)   & 11.365  &  236.0(12) \\
\hline
\end{tabular}
\end{center}
\caption{Monopole density in lattice units (fourth column) and
  physical units (last column) for different lattice sizes 
and inverse gauge couplings. In the fifth column we report the
  corresponding values of $T/T_c$. In order to determine the physical
scale in the last column, we have assumed $\sqrt{\sigma} = 430$ MeV;
a star indicates data obtained with a looser gauge fixing criterion 
$\omega = 10^{-3}$ ($\omega = 10^{-8}$ has been used in all other cases).
}
\label{denstab}
\end{table}

\begin{table}
\begin{center}
\begin{tabular}{|c|c|c|}
\hline
$L_s^3\times L_t$  &  Landau+$MAG$      & $MAG$ \\
\hline  
$40^3\times 7$ &     $1.56013(6)$  & $1.55898(7)$    \\ 
\hline
$32^3\times 5$ &     $1.53470(13)$ & $1.53354(14)$   \\ 
\hline
$32^3\times 6$  &     $1.53438(16)$ & $1.53335(17)$  \\
\hline
$32^3\times 7$ &     $1.53420(7)$  & $1.53313(6)$    \\
\hline
\end{tabular}
\caption{Average normalized maximum reached for the MAG functional, 
Eq.~(\ref{maxfun}), during the gauge fixing procedure, with and
without performing a previous gauge fixing to Landau gauge.
}
\label{MAGfunctab}
\end{center}
\end{table}

\begin{thebibliography}{9}


\bibitem{thooft75}   G. 't Hooft, in ``High Energy Physics'', EPS
  International Conference, Palermo 1975, ed. A. Zichichi.
\bibitem{mandelstam} S.~Mandelstam, Phys. Rept. {\bf 23}, 245 (1976).
\bibitem{parisi} G.~Parisi, Phys. Lett. {\bf B60}, 93 (1975).
\bibitem{superI-II}
A.~Di Giacomo, B.~Lucini, L.~Montesi, G.~Paffuti,
Phys.\ Rev.\ {\bf D 61}, 034503 (2000) 
[arXiv:hep-lat/9906024];
Phys.\ Rev.\ {\bf D 61}, 034504 (2000) 
[arXiv:hep-lat/9906025];
\bibitem{superIII}
J.~M.~Carmona, M.~D'Elia, A.~Di Giacomo, B.~Lucini, G.~Paffuti,
Phys.\ Rev.\ {\bf D 64}, 114507 (2001) 
[arXiv:hep-lat/0103005];
\bibitem{superfull}
J.~M.~Carmona, M.~D'Elia, L.~Del Debbio, A.~Di Giacomo, B.~Lucini, G.~Paffuti,
Phys.\ Rev.\ {\bf D 66}, 011503 (2002) 
[arXiv:hep-lat/0205025];
\bibitem{superIV}
M.~D'Elia, A.~Di Giacomo, B.~Lucini, G.~Paffuti, C.~Pica, 
Phys.\ Rev.\ {\bf D 71}, 114502 (2005) 
[arXiv:hep-lat/0503035].
\bibitem{moscow}
M.N.~Chernodub, M.I.~Polikarpov and A.I.~Veselov,
Phys.\ Lett.\ {\bf  B399}, 267 (1997).
\bibitem{bari}
P.~Cea and L.~Cosmai,
JHEP {\bf 0111}, 064 (2001); P.~Cea, L.~Cosmai and M.~D'Elia,
JHEP {\bf 0402}, 018 (2004). 
\bibitem{kortals}
C.~P.~Korthals~Altes,
hep-ph/0607154.
\bibitem{shuryak}
J.~Liao, E.~Shuryak,
Phys.~Rev.~{\bf C 75} 054907 (2007) [hep-ph/0611131].
\bibitem{shuryak1} 
J.~Liao, E.~Shuryak, arXiv:0706.4465 [hep-ph].
\bibitem{chezak}
M.~N.~Chernodub and V.~I.~Zakharov,
Phys. Rev. Lett. {\bf 98}, 082002 (2007) 
[arXiv:hep-ph/0611228]; 
M.~N.~Chernodub and V.~I.~Zakharov,
arXiv:hep-ph/0702245.
\bibitem{chelatt}
M.~N.~Chernodub, K.~Ishiguro, A.~Nakamura, T.~Sekido, T.~Suzuki and V.~I.~Zakharov,
arXiv:0710.2547 [hep-lat].
\bibitem{bornya92}
V.G.~Bornyakov, V.K.~Mitrjushkin and M.~Muller-Preussker
Phys. Lett. {\bf B284}, 99 (1992).
\bibitem{ejiri}
S.~Ejiri,
Phys. Lett. {\bf B376}, 163 (1996)
[arXiv:hep-lat/9510027].
\bibitem{degrand}
A.~De~Grand, D.~Toussaint, 
Phys.~Rev.~ {\bf D 22} 2478 (1980).
\bibitem{karsch}
J.~Engels, F.~Karsch, K.~Redlich, 
Nucl.\ Phys.\ {\bf B 435}, 295 (1995)
[arXiv:hep-lat/9408009].
\bibitem{karsch2}
J.~Fingberg, U.~Heller, F.~Karsch,
Nucl.\ Phys.\ {\bf B 392}, 493 (1992) 
[arXiv:hep-lat/9208012].
\bibitem{cosmai}
P.~Cea, L.~Cosmai 
Phys.~Rev.~\textbf{D 52}:5152 (1995) [hep-lat/9504008].
\bibitem{giovannangeli}
P.~Giovannangeli and C.~P.~Korthals~Altes, 
Nucl.\ Phys.\ {\bf B 608}, 203 (2001)
[arXiv:hep-ph/0102022].
\bibitem{kovacs}
T.~G.~Kovacs, E.~T.~Tomboulis
Phys.~Lett.~\textbf{B463}, 104 (1999) [hep-lat/9905029]. 
\bibitem{bornyakov}
G.~S.~Bali, V.~Bornyakov, M.~Mueller-Preussker, K.~Schilling, 
Phys.~Rev.~{\bf D 54} 2863 (1996) [hep-lat/9603012];
V.~G.~Bornyakov, D.~A.~Komarov, M.~I.~Polikarpov,
Phys.~Lett.~{\bf B497} 151 (2001).
\bibitem{greensite}
M.~Faber, J.~Greensite, S.~Olejnik
Phys.~Rev.~{\bf D 64}, 034511 (2001)  [hep-lat/0103030];
M.~Faber, J.~Greensite, S.~Olejnik, D.~Yamada,
JHEP {\bf 9912}, 012 (1999).
\bibitem{su3gribov}
A.~O~Cais, W.~Kamleh, K.~Langfeld, B.~Lasscock, D.~ Leinweber,
L.~von~Smekal,
arXiv:0710.2958
\bibitem{zak2}
V.~I.~Zakharov,
  Braz.\ J.\ Phys.\  {\bf 37}, 165 (2007)
  [arXiv:hep-ph/0612342].
\bibitem{bornya02}
V.~G.~Bornyakov, D.~A.~Komarov, M.~I.~Polikarpov and A.~I.~Veselov,
  arXiv:hep-lat/0210047.
\end{thebibliography}
\end{document}